# A Novel Study on Intelligent Methods and Explainable AI for Dynamic Malware Analysis.

RICHA DASILA, Indian Institute of Information Technology Allahabad, India.
VATSALA UPADHYAY, Indian Institute of Information Technology Allahabad, India.
PROF. SAMO BOBEK, University of Maribor, Slovenia
DR. ABHISHEK VAISH[*], Indian Institute of Information Technology Allahabad, India.

[*]Dr. Abhishek Vaish, abhishek@iiita.ac.in

Authors' Contact Information: Richa Dasila, Indian Institute of Information Technology Allahabad, Prayagraj, Uttar Pradesh, India., mcl2022011@iiita.ac.in;

Vatsala Upadhyay, Indian Institute of Information Technology Allahabad, Prayagraj, India., vatsau01@gmail.com;

Prof. Samo Bobek, University of Maribor, Maribor, Slovenia, samo.bobek@um.si;

Dr. Abhishek Vaish, Indian Institute of Information Technology Allahabad, Prayagraj, Uttar Pradesh, India., abhishek@iiita.ac.in.

Deep learning models are one of the security strategies, trained on extensive datasets, and play a critical role in detecting and responding to these threats by recognizing complex patterns in malicious code. However, the opaque nature of these models—often described as "black boxes"— makes their decision-making processes difficult to understand, even for their creators. This lack of transparency can undermine trust in their outputs. This research addresses these challenges by integrating Explainable AI (XAI) techniques to enhance the interpretability and trustworthiness of malware detection models. In this research, the use of Multi-Layer Perceptrons (MLP) for dynamic malware analysis has been considered, a less explored area, and its efficacy in detecting Metamorphic Malware and further the effectiveness and transparency of MLPs, CNNs, RNNs, and CNN-LSTM models in malware classification, evaluating these models through the lens of Explainable AI (XAI). This comprehensive approach aims to demystify the internal workings of deep learning models, promoting a better understanding and trust in their predictive capabilities in cybersecurity contexts. Such in-depth analysis and implementation haven't been done to the best of our knowledge.

## 1 Introduction

Malware is software that intrudes into the system without the user's knowledge, intending to harm the system's resources. There are various types of malware within the malware family, and each poses distinct threats, including viruses, worms, trojans, ransomware, rootkits, adware, and spyware (Rieck et al., 2008; Alaeiyan et al., 2019; Singh and Singh, 2019). In malware analysis, methodologies are divided into static and dynamic analysis (Shijo and Salim, 2015; Damodaran et al., 2017). Static analysis involves examining binaries without executing them, using techniques such as file fingerprinting, AVS scanning, packer detection, and disassembling to provide a comprehensive analysis. However, it struggles with obfuscated or encrypted malware, as it cannot reveal runtime behaviors. On the other hand, dynamic analysis entails executing the malware in a controlled environment to observe its behavior. This approach utilizes tools like Filemon and Norman to analyze different defined paths during execution and observe runtime behaviors. Dynamic analysis is particularly effective against obfuscated malware, including metamorphic variants that employ obfuscation techniques like dead code insertion, register re-assignment, subroutine re-ordering, instruction substitution, and code transposition to change their code with every scan, thereby evading anti-virus scanners and enabling zero-day attacks (Or-Meir et al., 2019; Bhadran and Kapadia, 2023; Willems et al., 2007). Dynamic analysis is crucial for tackling these sophisticated and elusive threats. Metamorphic malware is an active research domain in the cybersecurity discipline (Jha et al., 2022).

In contrast to traditional malware, dynamic malware can dynamically alter its code, behavior, and signature in real time to elude detection and analysis (Bayer et al., 2010). Employing techniques such as polymorphism (which mutates static binary code through encryption keys) and metamorphism (reprogramming itself by translating its code and rewriting subsequent copies with the assistance of a metamorphic engine), dynamic malware actively seeks to evade detection systems and antivirus software. Its adaptive nature allows it to modify behavior based on its environment, making signature-based antivirus solutions often ineffective against dynamic malware. The efficiency of machine learning techniques in addressing the increasing complexity of dynamic malware is supported by studies (Mangialardo and Duarte, 2015; Khalid et al., 2023; Moon et al., 2021; Akhtar and Feng, 2023). These techniques learn from new malware samples, adapting to emerging threats and continuously improving detection capabilities.

Machine learning (ML) has been extensively applied in malware detection, utilizing algorithms designed to predict outcomes without explicit programming (Aslan and Samet, 2020). ML aims to convert input data into acceptable value ranges via statistical analysis, enabling classification, regression, and clustering operations. Notable ML algorithms, such as Bayesian networks (BN), naive Bayes (NB), C4.5 (J48), logistic model trees (LMT), random forests (RF), k nearest neighbors (KNN), simple logistic regression (SLR), support vector machines (SVM), and sequential minimal optimization (SMO), have long been used in malware detection (Kumar et al., 2023). Each algorithm has distinct advantages and limitations, influenced by data distribution and feature interdependencies. Deep learning models, including multilayer perceptrons (MLPs), convolutional neural networks (CNNs), recurrent neural networks (RNNs), long short-term memory networks (LSTMs), and autoencoders, have demonstrated exceptional performance in processing complex, high-dimensional data, further enhancing ML's capability in advanced malware detection systems (Bensaoud et al., 2024).

However, several challenges, such as the interpretability and trust of these models, affect their effectiveness. ML/DL models (Naseer et al., 2021; Hi-taj et al., 2024), especially those based on deep learning, are often seen as "black boxes" because it is challenging to understand their decision-making processes. This opacity can lead to mistrust among cybersecurity professionals who rely on these tools. To address this issue, it is crucial to integrate methods that enhance the transparency of these models.

By emphasizing the features that significantly influence predictions—such as identifying key characteristics that lead a model to classify a file as malware—the trust and reliability in ML/DL models can be significantly improved. These approaches make the models more user-friendly and accountable, ensuring they can be effectively used in sensitive security applications.

The lack of interpretability in traditional ML models hinders their acceptance among cybersecurity professionals, as they require explicit artifacts to take informed actions. Moreover, conventional ML models often struggle to adapt to new, evolving threats and manage the high dimensionality of malware data, resulting in false positives and negatives. XAI can address these challenges by providing transparency in the decision-making process of ML models, enabling security experts to understand, trust, and respond effectively to the insights generated. XAI techniques can highlight the features most indicative of malicious behavior, refining feature extraction processes and improving model adaptability to novel threats. By enhancing interpretability and reliability, XAI can significantly boost the effectiveness of malware detection and analysis. This is where Explainable Artificial Intelligence comes into the picture. With the help of techniques like SHAP (Shapley Additive Explanations) and LIME (Local Interpretable Model-agnostic Explanations), it becomes possible to offer detailed insights into the reasoning behind model predictions (Alenezi and Ludwig, 2021).

SHAP values elucidate how much each feature contributes to the outcome, whether pushing the model's prediction higher or lower. This is particularly useful in understanding complex models in a high-stakes domain like cybersecurity, where knowing which bytes, permissions, or API calls influence the decision can be critical. On the other hand, LIME provides local explanations, allowing users to see how changes in input features affect predictions on a case-by-case basis (Saqib et al., 2024). These techniques make ML/DL models transparent and allow for fine-tuning and error correction by pinpointing inaccuracies in the features that lead to false predictions. Thus, these approaches significantly enhance the trustworthiness and efficacy of ML/DL applications in malware analysis.

The novelty of this work is that it uses the concept of Explainable AI in cybersecurity with a specific use case of metamorphic malware and analyzes the performance of MLP in detection. In this research, we deploy four Deep learning models: CNN, RNN, CNN-LSTM, and MLP. MLP has not been explored yet in detecting malware that interacts with the system through API calls; that is, the malware that requires dynamic analysis. We analyze the factors that affect a model's performance, which is divided into various sections: Section 2 delves into related work, establishes a foundation for malware classification and Explainable AI (XAI), and identifies performances from both the operational and XAI perspectives using LIME and SHAP.

This paper is systematically organized to provide a comprehensive understanding of our research, divided into two main parts: The first advocates the use of the Multi-Layer Perceptron (MLP) as a potential model for malware detection as a base method for detection of Malware, while the second analyzes four different models from both Operational i.e. to run series of test with different input and evaluate the performance of different models for e,g MLP, CNN, RNN, CNN-LSTM to present a cohesive understanding and XAI perspectives to understand the insight of the models for a transparent and trustworthy understanding of the models. The paper is divided into sections: Section 2 delves into related work, establishes a foundation for malware classification and Explainable AI (XAI), and identifies research gaps. Section 3 covers the research methodology, the models used in our work, their brief explanations, the dataset description, and the brief working of SHAP and LIME. Section 4 discusses the results and contributions of the proposed work and details the experiments, which are divided into operational and Explainable parts. The operational part consists of various experiments performed on the MLP model and a comparative analysis with existing models. The analysis of the Explainable AI part involves the analysis of all four models using LIME and SHAP. Section 5 is the paper's conclusion, summarizing our findings and limitations.

## 2 Related Work

Though Machine Learning (ML) has demonstrated high accuracy in detecting malware, its real-world performance often falls short of expectations, primarily because the distribution of goodware significantly outnumbers malware in practical scenarios—less than 1% of all executables are reported as malware (web). This discrepancy leads to a performance drop when models transition from controlled environments to deployment, compromising analysts' trust. Furthermore, as stated in (Smith et al., 2020a), many training datasets are inherently biased, resulting in overperformance during testing. The datasets used in malware analysis typically include only detected samples, omitting undetected malware and thus limiting feature representation. ML models extract pre-determined features from malware, but this approach may miss novel or evolving malicious behaviors. The inherent bias in the dataset, where only samples flagged by tools are labeled as malware and others are discarded, exacerbates the issue. Studies have shown that a balanced training set (1:1 ratio) provides a suitable evaluation framework, but only when the imbalance exceeds a 1:8 ratio do performance metrics reflect real-world effectiveness (Illes, 2022). These factors undermine the reliability of ML models in large-scale deployments for malware detection, raising critical questions about their pr. The RNN model had an accuracy of 79%, and the ANN model—which uses a multi-layer perceptron (MLP)—showed 91% neural network (RNN) and artificial neural network (ANN) models for ransomware detection and classification. The dataset used for the experiment included 427 files that correspond to nine different families of ransomware (bimtan, cerber v1, cerber v2, fsysna, telsacrypt, xorist, yakes, zerberV1, and zerber v2). A two-step methodology was proposed, and the research first classifies programs as malware or goodware based on system calls. Then, it uses CNN, RNN, and ANN models for binary classification and categorizing the nine kinds of ransomware. The results indicate that the CNN model had an accuracy of 94%, the RNN model had an accuracy of 79%, and the ANN model—which uses a multi-layer perceptron (MLP)—showed 91% accuracy for identifying ransomware families. Such high accuracy could be attributed to the small dataset size, but MLP showed 100% accuracy for binary classification. An extensive analysis uses the Malign data set (Ben Abdel Ouahab et al., 2022), consisting of 9,339 malware pictures from 25 families. The study uses malware visualization techniques and MLP to find patterns in grayscale samples and assign photos to different classes based on those patterns. The suggested classification model uses the GIST descriptor as an image feature, which is then integrated into the input layer. Two experimental setups were investigated: one with one hidden layer and the other with two hidden layers. The model with 30 neurons in layer 1 and 100 neurons in layer 2 attained a maximum accuracy of 0.976. This accuracy was higher than that of previous models, which used gray-scale images, and motivated us to analyze how and to what extent MLP has been used previously.

(Singh and Singh, 2021), The paper discusses the application of behavior-based methods for malware detection, focusing on runtime characteristics. A data set consisting of 3231 malware files was collected from different sources. The research involved creating a dynamic analysis setup using Cuckoo Sandbox, extracting eight distinct features: file operations, registry files, network files, API calls, DLL files, Processes, Mutex operations, and Import functions. These features were then utilized in a binary classification model, employing a Multi-Layer Perceptron (MLP), resulting in an accuracy rate of 99.2%. The study did not incorporate any specific algorithmic approach for the malware classifier training. Although artificial intelligence-based approaches for detecting and defending against cyber attacks are more advanced and efficient than conventional signature-based and rule-based strategies, most ML-based and DL-based techniques are deployed as "black boxes." AI techniques, particularly ML and DL algorithms, show impressive performance in intrusion detection, spam email filtering, botnet detection, fraud detection, and malicious application identification (Sahakyan et al., 2021). However, they can still make costly errors and often sacrifice interpretability for higher accuracy, leading to more complex models (Mallick et al., 2023). Regulations like the European Union's General Data Protection Regulation (GDPR) highlight the need for transparency in AI decisions, especially those impacting individuals negatively (Goodman and Flaxman, 2017). To build trust, AI in cybersecurity must be transparent and interpretable, with XAI techniques already being implemented in various fields, such as healthcare, natural language processing, and financial services (Nazar et al., 2021). With the help of Explainable AI (XAI), we can better comprehend and trust the results produced by machine learning algorithms (Renftle et al., 2022; Madanu et al., 2022), aligning ML solutions more effectively with malware detection—a critical issue in our established network. The literature review presented above underscores the effectiveness of the Multi-Layer Perceptron (MLP) in malware detection. These studies collectively provide a strong foundation for using MLP for dynamic malware detection despite not being widely experimented with in this context. Therefore, our research will employ MLP, building on the evidence presented to further explore its potential in dynamic malware detection and to use the concept of Explainable AI because the lack of transparency and interpretability can erode confidence in these models, especially as cyber-attacks become more diverse and complex. Therefore, it is crucial to apply Explainable AI (XAI) to develop cybersecurity models that are both accurate and understandable, allowing users to comprehend, trust, and manage these systems effectively (Mathews, 2019).

2.1 Research Gaps
After the detailed LR presented above, the following research gaps have been observed:

2.2

Integrating MLP models with API calls of malware is an unexplored area in research. This study aims to fill this gap by examining how MLP works with API integration as a base model for Malware detection and analyzing other deep learning models. We will set up an environment to test this and compare it with other algorithms. This investigation will help understand the feasibility, benefits, and challenges, potentially leading to new applications and better model usability.

2.3

There is a lack of behavioral datasets present for training models. Data in Malware Analysis (MA) significantly differs from other domains as it lacks proximity relationships, continuity, and ordinality (Ge et al., 2021; Smith et al., 2018). This disparity means that datasets often do not accurately represent real-world problems, affecting the robustness of models trained on such data. A high false negative rate is also expected when the dataset includes a broader set of goodware. Datasets in this field are inherently biased, leading to model over-performance. Furthermore, samples that do not have labels are often discarded, which limits the research from a versatile perspective. Since malware is behavioral in nature, current machine-learning techniques struggle to capture these behaviors, making malware analysis deficient with the available datasets. Moreover, current feature extraction and machine learning techniques optimized for signal processing are inadequate for addressing malware behaviors.

2.4

There is a significant research gap in the practical implementation and analysis of XAI and ML/DL techniques in malware analysis (Smith et al., 2020b; Sommer and Paxson, 2010). The current literature is largely comprised of surveys that are found to be very middle-ended, and research papers often focus narrowly on one specific problem, dataset, or model without providing a comprehensive analysis. As a result, there is a lack of studies that offer a thorough, practical, and reliable implementation of XAI in real-world malware detection scenarios. This gap highlights the need for more complete research that not only addresses key questions but also sequentially and methodically carries out the analysis, providing detailed insights into the effectiveness of XAI in enhancing the transparency, trustworthiness, and performance of ML/DL models in malware analysis.

**3 Research Methodology**

We have used an experimental research method for the proposed research, shown in Fig. 1. This section will explain the models used, the dataset, the number of experiments, and system requirements. This research methodology is divided into two parts: Part 1 consists of the proposed MLP method and its subparts, as shown in Figure 3, and Part 2 consists of our XAI approaches.

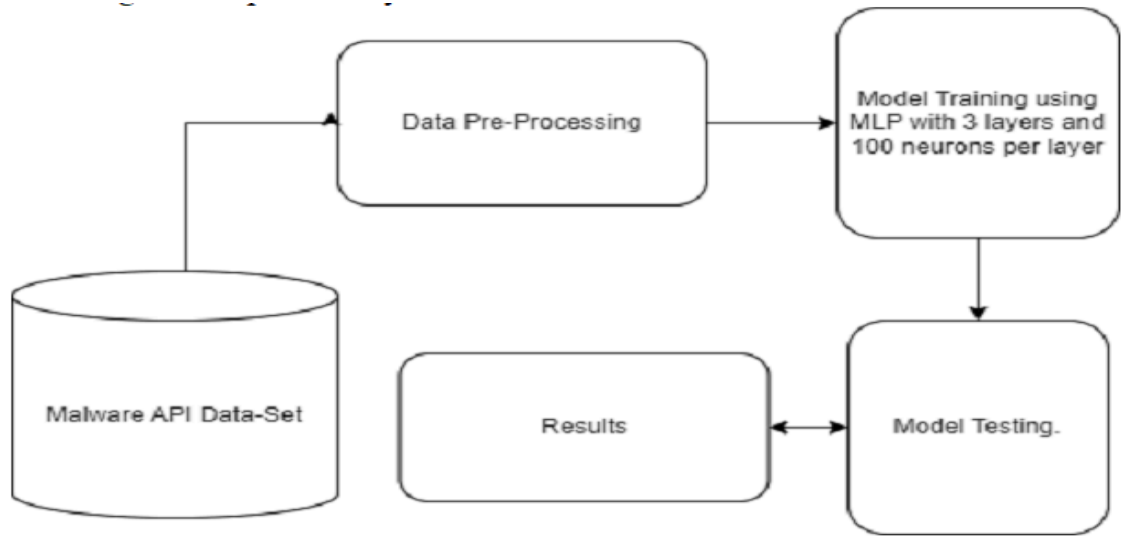

Fig. 1. The research design

3.1 Data-Set - Part 1

To analyze and understand the behavior of dynamic malware, i.e., malware that changes its code or actions to avoid detection, we must run the malware in a controlled environment. This allows us to observe its actions in real-time. Among the various behaviors we can monitor, the sequence of API calls made by the malware is particularly critical. This sequence forms a behavioral fingerprint, revealing how the malware interacts with the system and what malicious actions it attempts to perform. We leverage the API-Sequence Dataset, which comprises 42,797 sequences of malware API calls and 1,079 sequences of goodware API calls for our research work.

The constructed dataset features 101 columns: the initial column is reserved for the MD5 hash of each sample, followed by 100 columns that map out the sequence of API calls for the sample, with these calls indexed from 0 to 306, reflecting the 307 unique API calls documented in the study. The entire data collection process spanned 3000 hours, generating around 50,000 Cuckoo JSON reports and amassing 1.5TB of unprocessed data. The Cuckoo Sandbox configuration employed for this dataset included an Intel Xeon D-1540 server with 8 cores and 16 threads, operating at 2.6GHz, with 64GB RAM and a 2TB SSD, running Ubuntu Server 16.04. Additionally, eight 32-bit Windows 7 Ultimate instances were run on VirtualBox in parallel as part of the Cuckoo analysis setup. Our research analyzes API call sequences as a crucial method for dynamic malware detection. This is grounded in the premise that malware, during execution, often displays behavior distinguishable from benign software through its interactions with the operating system, typically done through API calls. It's important to note that our approach does not consider intelligent malware capable of deceiving sandboxing environments. For our experiments, we used a setup known as an n1-highmem-2 instance. This setup included two virtual CPUs (vCPUs) running at a speed of 2.2 gigahertz (GHz), and we used the T4 GPU for all models, which provided a solid foundation for our analysis and experimental tasks. To ensure the smooth running of programs and effective handling of data, the system came with 13 gigabytes of RAM. Additionally, we allocated a generous 100 gigabytes of free disk space. Our analytical process unfolded in four distinct stages, as shown in Figure 3. Initially, we carefully selected the dataset tailored for dynamic analysis, settling on the previously identified malware API dataset.

3.2 Model used for Explainable AI (XAI) - Part 2:

Recurrent Neural Networks (RNNs) are deep learning models designed for processing sequential data by leveraging their ability to maintain temporal dependencies. Our RNN model uses an embedding layer to convert input indices into 100-dimensional dense vectors. The recurrent layer employs Long Short-Term Memory (LSTM) cells, which mitigate the vanishing gradient problem and improve long-range dependency capture. The LSTM layer has 50 hidden units, operates bidirectionally to capture past and future contexts, and includes dropout for regularization.

- Embedding Layer: Transforms input indices into 100-dimensional vectors.
- LSTM Layer: Processes embeddings with 50 hidden units, bidirectional processing, and 0.2 dropouts.

- Fully Connected Classifier: Two linear layers, with ReLU activation and a sigmoid output for binary classification.

This architecture effectively captures temporal features, making it suitable for sequential data tasks like malware detection.

Convolutional Neural Networks (CNNs) are deep learning models designed to process grid-like data, such as images or sequences, by capturing spatial hierarchies of features. CNNs use convolutional, pooling, and fully connected layers to extract and reduce features efficiently. Our model transforms input indices into 100-dimensional vectors using an embedding layer. This is followed by three convolutional blocks, each consisting of a convolutional layer, ReLU activation, batch normalization, dropout, and max pooling. These layers extract high-level features and reduce dimensionality. An adaptive average pooling layer further reduces the feature map size.

- Embedding Layer: Transforms input indices into 100-dimensional vectors.

- Convolutional Layers: Three convolutional blocks with convolutional layers, ReLU activation, batch normalization, dropout, and max pooling.
- Fully Connected Layers: Linear layers with ReLU and sigmoid activations to produce a probability score for binary classification.

This architecture is well-suited for tasks like malware detection.

The CNN-LSTM architecture combines Convolutional Neural Networks (CNNs) and Long Short-Term Memory (LSTM) networks to leverage the strengths of both models. This hybrid approach is particularly effective for tasks requiring spatial feature extraction and temporal sequence modeling, such as time-series analysis and sequential data processing. In our model, the embedding layer first transforms input indices into 100-dimensional dense vectors. This is followed by a batch normalization layer to stabilize and accelerate the training process. The convolutional layer then applies 32 filters with a kernel size of 1D convolution to capture local spatial patterns in the data, followed by a max pooling layer to reduce the dimensionality and retain the most salient features. The LSTM layer processes these extracted features with 512 hidden units to capture long-term dependencies and temporal patterns within the sequential data. Finally, a dense (fully connected) layer with a sigmoid activation function produces the output probability for binary classification. The parameter values have been illustrated in Fig. 2.

- Embedding Layer: Transforms input indices into 100-dimensional dense vectors.
- Batch Normalization: Normalizes the output of the embedding layer to improve training stability.
- Convolutional Layer: Applies 32 filters of 1D convolution to capture spatial patterns.
- Max Pooling Layer: Reduces the dimensionality by selecting the maximum value from each filter.
- LSTM Layer: Processes the pooled features to capture temporal dependencies with 512 hidden units.
- Dense Layer: Outputs the final probability score for binary classification using a sigmoid activation function.

```
Layer (type)                    Output Shape              Param #
=================================================================
layer_embedding (Embedding      (None, 100, 8)            2456
)

batch_normalization (Batch      (None, 100, 8)            32
Normalization)

conv1d (Conv1D)                 (None, 100, 32)           2336

max_pooling1d (MaxPooling1      (None, 50, 32)            0
D)

lstm (LSTM)                     (None, 512)               1116160

dense (Dense)                   (None, 1)                 513

=================================================================
Total params: 1121497 (4.28 MB)
Trainable params: 1121481 (4.28 MB)
Non-trainable params: 16 (64.00 Byte)
```

Fig. 2. The Screenshot of the CNN Architecture along with the Values.

The mathematical formulation for the CNN-LSTM model is as follows:

$$E = \text{Embedding}(X) \quad (1)$$

$$Z = \text{ReLU}(\text{Conv1D}(E, W) + b) \quad (2)$$

$$P = \text{MaxPool}(Z) \quad (3)$$

$$(H_t, C_t) = \text{LSTM}(P, H_{t-1}, C_{t-1}) \quad (4)$$

$$Y = \sigma(W_f \cdot H_t + b_f) \quad (5)$$

This model architecture effectively captures spatial and temporal features, making it well-suited for complex sequential data tasks such as malware detection.

Multi-layer perceptron (MLP) is a feed-forward artificial neural network (ANN). ANN is a computer model inspired by the biological workings of the human brain. ANN has three layers: the input layer, the hidden layer, and the output layer. MLP is an ANN in which all nodes are connected to all other nodes of different layers. Each node passes its value to upcoming nodes only in the forward direction. To increase the accuracy of the model, MLP uses a backpropagation algorithm. While a Perceptron excels in classifying linearly separable data due to its linear nature, the Multi-layer Perceptron (MLP) takes a more sophisticated approach. By leveraging the power of multiple neurons organized across at least three layers, namely the input layer, which receives and processes features; the successive hidden layers, responsible for further processing and distributing features; and finally, the output layer of the perceptron, which produces the ultimate output based on the input from the last hidden layer. Illustrated in Fig.3 is a depiction of a three-layer MLP with three input features and 4 neurons per layer, showcasing the intricate network that enables MLPs to approximate continuous functions effectively. Backpropagation is the learning mechanism that MLP uses to calculate the gradient of the loss function. It is used to increase the accuracy of the output by reducing the error in the predicted output and the actual output. Another important aspect is the choice of activation functions in hidden layers. It determines the output of a neuron for the given input. It introduces non-linearity, enabling the network to learn complex relations. Activation functions for MLP are the Rectified Linear Unit(ReLU) Function, Logistic Sigmoid function, and Hyperbolic Tan function(tanh). Corresponding equations of each activation function are given in Table 1.

Perceptron MLPs, or Multilayer Perceptrons, are fundamentally composed of perceptrons, initially conceptualized by Frank Rosenblatt as a supervised learning algorithm. They receive *n* features from inputs ($x_1, x_2, \ldots, x_n$), and each feature is associated with a weight ($w_1, w_2, \ldots, w_n$). The perceptron calculates the weighted sum of the input features with the formula:

$$\text{Weighted Sum} = w_1x_1 + w_2x_2 + \cdots + w_nx_n = \sum_{i=1}^{n} w_i x_i \quad (6)$$

A bias term *b* is also added to the weighted sum, which acts like an intercept in the linear equation and allows the perceptron to fit the data. The adjusted sum is computed as:

$$\text{Adjusted Sum} = \sum_{i=1}^{n} w_i x_i + b \quad (7) \quad\quad (7)$$

It uses a simple activation function, which outputs 1 if the adjusted sum is greater than or equal to 0 and 0 otherwise. The activation function can be represented as:

$$\text{Activation Function} = \begin{cases} 1 & \text{if Adjusted Sum} \geq 0 \\ 0 & \text{otherwise} \end{cases} \quad (8)$$

Local Interpretable Model-agnostic Explanations (LIME) is a technique designed to improve the interpretability of complex, black-box machine learning models. LIME provides insights into model predictions by approximating the black-box model with an interpretable model locally around the prediction of interest.

LIME's primary objective is to explain any classifier's predictions in an interpretable and faithful manner. It does so by perturbing the input data and observing the prediction changes. This allows it to create a local surrogate model—usually a simple, interpretable model like linear regression or a decision tree—that approximates the behavior of the complex model in the vicinity of the prediction. By fitting a simple model to the perturbed data, LIME provides insights into which features are most influential in the black-box model's prediction. This approach is model-agnostic, meaning it can be applied to any machine learning model regardless of its complexity. LIME has been widely adopted for its ability to enhance the transparency of machine learning models, making it easier for users to understand and trust the predictions. This is especially important in sensitive applications such as healthcare, finance, and cybersecurity, where understanding the rationale behind a prediction is crucial. SHapley Additive explanations (SHAP) is a unified framework for interpreting predictions of machine learning models. SHAP leverages concepts from cooperative game theory, particularly the Shapley value, to attribute the contribution of each feature to the prediction of a model. This approach provides a consistent and theoretically sound method to explain individual predictions, making it one of the most robust tools for model interpretability.

The Shapley value is a solution concept from cooperative game theory that ensures fair distribution of payouts among players based on their contributions. In machine learning, features are treated as players; the prediction is the payout. The SHAP value for a feature represents its contribution to the difference between the actual and average predictions, considering all possible combinations of features. The calculation of SHAP values involves the following steps:

• Coalition Formation: Create all possible subsets (coalitions) of features.
• Marginal Contribution: For each coalition, compute the change in the model's prediction when the feature is added to the coalition.
• Average Marginal Contribution: The SHAP value for a feature is the average of its marginal contributions across all coalitions.

Mathematically, the SHAP value for a feature $i$ is given by:

$$\phi_i = \sum_{S \subseteq N \setminus \{i\}} \frac{|S|!(|N| - |S| - 1)!}{|N|!} [f(S \cup \{i\}) - f] \tag{9}$$

where:

- $N$ is the set of all features,
- $S$ is a subset of $N$ not containing feature $i$,
- $f(S)$ is the model's prediction for the feature subset $S$.

SHAP values provide several advantages:

• Consistency: The sum of SHAP values for all features equals the difference between the model's prediction and the average prediction.
• Local Accuracy: The SHAP values for a specific instance add to the model's prediction. • Global Interpretability: Aggregating SHAP values across multiple instances offers insights into the overall importance of features.

By applying SHAP to our models, we gain detailed insights into how each feature influences individual predictions and the model. This enhances the interpretability and transparency of our machine-learning models, providing valuable information for model validation and debugging.

## 4 Results

The result section is presented in two parts. The first part is the use of MLP and its efficacy in detecting Metamorphic Malware; this is the outcome of research gap 2.2. The working of the MLP is presented, followed by the performance analysis using well-established performance metrics along with various test cases to check the performance under data variability, which is the outcome of research gap 2.3, and to observe the outcome, a comparison of the performance using other deep learning models is also presented to present a cohesive analysis of the proposed work. The second part of this section is the understanding of the dynamics of MLP and another deep learning method in the light of Explainable AI (XAI); the motivation for this part is the outcome of research gap 2.4.

Table 1. Definition of Activation Functions

| Activation Function | Mathematical Expression |
|---|---|
| Sigmoid | $\sigma(x) = \frac{1}{1+e^{-x}}$ |
| Tanh | $\tanh(x) = \frac{e^x - e^{-x}}{e^x + e^{-x}}$ |
| ReLU | $f(x) = \max(0, x)$ |

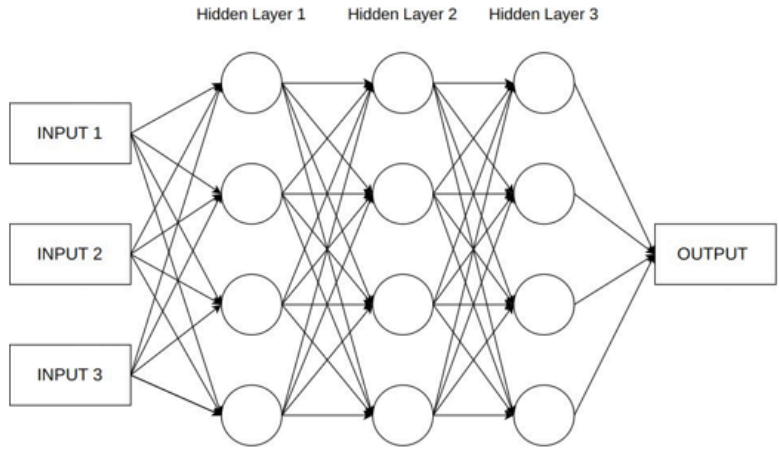

Fig. 3. A three-layer MLP with three hidden layers.

4.1 MLP Analysis

The algorithm for the Multi-Layer Perceptron is as follows:

(1) Forward Propagation:
   (a) Let *X* be the input vector.
   (b) For each layer *l* from 1 to *L* (where *L* is the number of layers):
      • Calculate the input to the layer:

$$Z^{(l)} = W^{(l)} A^{(l-1)} + b^{(l)}$$

where $A^{(0)} = X$, $W^{(l)}$ and $b^{(l)}$ are the weights and biases for layer $l$, and $A^{(l-1)}$ is the activation from the previous layer.

- Apply the activation function:
$$A^{(l)} = g^{(l)}(Z^{(l)})$$
where $g^{(l)}$ is the activation function for layer $l$.

(2) Calculate Output Error:
 - Let $Y$ be the actual output vector.
 - Calculate the error:
$$E = YA^{(L)}$$
where $A^{(L)}$ is the activation of the output layer.

(3) Backpropagation:
 (a) For each layer $l$ from $L$ to 1:
 - Calculate the gradient of the error concerning the weights:
$$\frac{\partial E}{\partial W^{(l)}}$$
 - Calculate the gradient of the error concerning the biases:
$$\frac{\partial E}{\partial b^{(l)}}$$
 - Update the weights and biases:
$$W^{(l)} = W^{(l)} - \eta \frac{\partial E}{\partial W^{(l)}}$$
$$b^{(l)} = b^{(l)} - \eta \frac{\partial E}{\partial b^{(l)}}$$
where $\eta$ is the learning rate.

(4) Repeat Until Convergence:
- Continue the process of forward propagation and backpropagation until the error $E$ is minimized.

*4.1.1 Performance of MLP for Malware Detection as a Reference Model.*

Given the binary nature of our sample classification (malware or benign), we employed a confusion matrix to showcase the outcomes depicted in Fig. 4. Our model demonstrated impressive performance, boasting a training score of 99.72% and a test score of 98.35%. The proposed model's efficiency is further underscored by an accuracy of 0.98, a precision reaching 0.9628, and an F1 score of 0.80, as in Table 2. These compelling results strongly advocate for the adoption of MLP in detecting dynamic malware, providing a clear and robust direction for future endeavors in this field.

| Metric | Value |
|---|---|
| Accuracy | 0.98 |
| Precision | 0.96 |
| F1 Score | 0.80 |

Table 2. Performance Metrics

*4.1.2 Effect of Neurons.* The growth of neurons does not exhibit a predictable pattern with an increase in the number of hidden layers. Instead, it appears to be more of a trial-and-error approach. We aim to avoid unnecessarily escalating

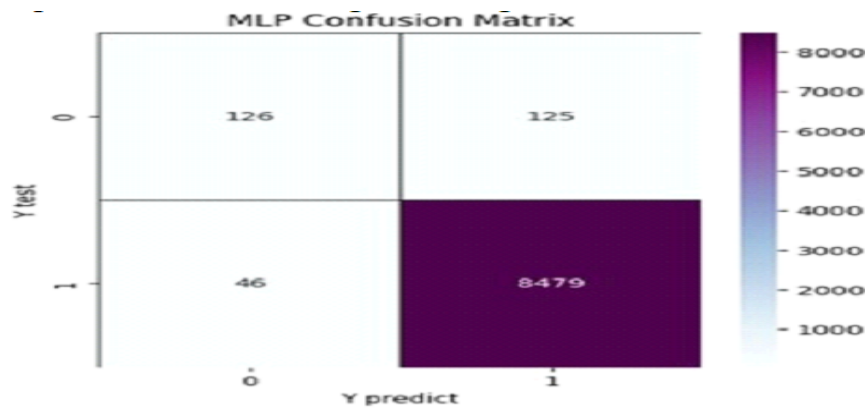

**Fig. 4. Screenshot of the Confusion Matrix of MLP**

| No. of layers | True Positive | False Positive | False Negative | True Negative | Avg. Precision Recall |
|---|---|---|---|---|---|
| 3 | 126 | 125 | 46 | 8479 | 0.99 |
| 4 | 137 | 114 | 51 | 8474 | 0.99 |
| 5 | 140 | 111 | 45 | 8480 | 0.99 |

Table 3. Performance metrics for different layer configurations

The complexity of our model, and there is no necessity to transition from a Multi-Layer Perceptron to a deep neural network, especially considering our research path. Maintaining a consistent number of neurons, we computed the average precision-recall values for each experimented model, as outlined in Table 3. Upon analysis, we observed that augmenting the number of layers has no discernible impact on the model's performance. As a result, we are content with our findings, leading us to assert that the complexity factor does not significantly influence the model's outcomes.

*4.1.3 Detailed Analysis using comparison of different models and the data variability concept.*

In this study, we focused on dynamic malware analysis. We selected three models that have been previously highlighted in the literature: Convolutional Neural Networks (CNN), Recurrent Neural Networks (RNN), and Support Vector Machines (SVM), as referenced in [1,2]. These models were trained using the selected dataset, adhering to the methodologies outlined in Fig. 1 and aligning with our experimental setup.

| Model | Accuracy | Precision | F1 Score |
|---|---|---|---|
| MLP | 0.98 | 0.96 | 0.88 |
| SVM | 0.97 | 0.94 | 0.96 |
| CNN | 0.96 | 0.98 | 0.97 |
| RNN | 0.96 | 0.97 | 0.97 |

Table 4. Model Performance Comparison

Following this comprehensive training phase, we conducted a detailed analysis of each model, evaluating them based on several key performance metrics: training time, testing time, accuracy, precision, and F1-score shown in Table 3. Our findings revealed that the MLP has demonstrated superior performance in terms of accuracy compared to the other models. Notably, the CNN model achieved the highest precision, recording a remarkable score of 0.98. Meanwhile, the

CNN model, along with the RNN model, excelled in the F1 score, achieving an impressive 0.97 (shown in Table 4).

In evaluating the performance of our models, while the CNN achieves a higher F1 score, we assert that the MLP demonstrates superior overall performance based on its higher accuracy. The F1 score, the harmonic mean of precision and recall, is particularly useful in cases of class imbalance, as it emphasizes the trade-offs between false positives and false negatives. Accuracy provides a broader measure of the model's ability to correctly classify malware and benign samples, which is critical for our application. Thus, while the CNN's F1 score may suggest better handling of edge cases, the MLP's higher accuracy indicates better generalization across the entire dataset, leading us to conclude that accuracy, rather than the F1 score, is the more appropriate metric for assessing overall model performance in this context. The dataset is unbalanced for this experiment.

We created a new dataset from the existing data to evaluate the model's performance on a balanced dataset. We extracted 1079 malware and 1079 benign samples and combined them using the pd.concat method. We utilized the random state parameter to ensure reproducibility, guaranteeing consistent randomness each time the code runs. Upon training the model on this balanced dataset, we observed an accuracy of 0.78. This demonstrates that the dataset's composition significantly affects model performance. Therefore, creating a dataset that closely mirrors real-world scenarios is essential to align our models with actual environments. We used SMOTE to overcome imbalances in our model. The advantages of using SMOTE include increased representation of the minority class, improved model performance, and more balanced training data. However, our model gave us an accuracy of 0.978, much higher than when we trained it on the balanced dataset curated in the previous section. This hints that our model may be overfitting due to the synthetic nature of the additional data. We manually attempted to address the class imbalance in our model by running the same script with different malware-to-benign ratios. For each ratio, we trained the model once randomly, then from top to bottom (using the top 60% of data for training and the rest for testing), and finally from bottom to top (using the bottom 60% of data for training and the rest for testing), discretizing the dataset in a ratio of 80:20, 60:40, 50:50 and finally 40:60 labeled as "legitimate" and "Forged" respectively as indicated in Table 5, We analyzed the behavior of our model and found that in most cases (as shown in Table 5), random training gave better accuracy. The graph obtained from this analysis is shown in Figure 5. It indicates that random training exhibited stable performance. In contrast, the change in the order of training, for example, the top-down ratio approach and bottom-up training approach, as explained earlier, depicts unstable performance with the value of accuracy highly dispersed, which is indicated in Table 5. Therefore, our model should be trained randomly. We used epochs to handle overfitting and applied random training as selected in the previous segment. We trained our model on the balanced dataset, initially giving an accuracy of 0.78. With a batch size of 150 and multiple epochs, we achieved an accuracy of 0.83. The advantages of using epochs include improved model convergence, better learning progression, and controlled overfitting. The model's performance improved with the use of epochs. After extensive training and testing on the MLP model, we found that MLP is a promising approach.

## 4.2 The Analysis of Malware Detection using SHAP and LIME (XAI)

Consequently, we move to the next part of our paper, which involves the analysis of deep learning models with XAI. We consider the best practices from our analysis to find solutions closer to real-world scenarios. Specifically, we use our MLP model with 3 layers of 100 neurons each and run it for multiple epochs with a batch size of 512 to minimize overfitting. The first one contained the exploration of MLP, an undiscovered path for API calls that are said to be the communication medium of dynamic malware analysis, and the second one contained XAI, which is further divided into operational and Explainable parts. In the first operational subpart, we analyze the models by various experiments on the dataset to note how the models behave. Then, we see each model in an Explainable way by applying SHAP and LIME. We analyze the results and try to provide insight into the AI model used in malware detection, trying to convert the black box model to a white box.

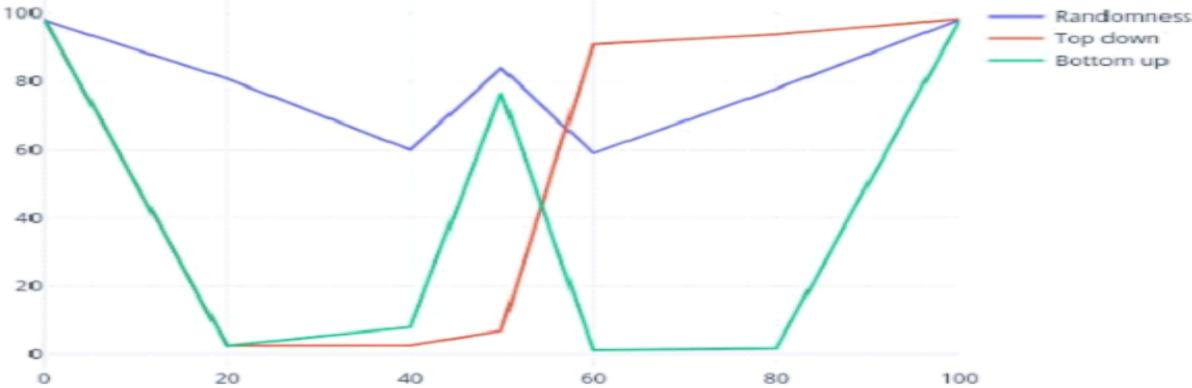

Fig. 5. Graph for Manual Experiments

| Legitimate | Forged | Training | Testing | Randomness | Accuracy |
| --- | --- | --- | --- | --- | --- |
| 100% (1-43,877) | 0 | 80% (random) | 20% (random) | Yes | 98.3% |
| 100% (1-43,877) | 0 | 80% | 20% (random) | No | 98.5% |
| 100% (1-43,877) | 0 | 80% (8776-4,8777) | 20% (1-8776) | No | 98.2% |
| 80% | 20% | 80% (random) | 20% (random) | Yes | 78% |
| 80% | 20% | 80% (1-35,101) | 20% | No | 94% |
| (8776-43,877) | | | (35,102-43,877) | | |
| 80% | 20% | 80% (random) | 20% (1-8776) | No | 1.7% |
| 60%(17,551-43,877) | 40% | 80% (random) | 20% | Yes | 59% |
| 60% | 40% | 80% (1-35,101) | 20% (35,102-43,877) | No | 91% |
| 60% | 40% | 80% (8776-48777) | 20% (1-8775) | No | 0.012% |
| 50% (1-26,503) | 50% (26,503-43,877) | 80% (random) | 20% (random) | Yes | 84% |
| 50% (1-26,503) | 50% (26,503-43,877) | 80% (1-35,101) | 20% (35,102-43,877) | No | 6.8% |
| 50% (1-26,503) | 50% (26,503-43,877) | 80% (8774-48777) | 20% (1-8777) | No | 76.7% |
| 50% (1-26,503) | 50% (26,503-43,877) | 80% (8774-48777) | 20% (1-8777) | No | 76.7% |
| 40% (21,940-43,877) | 60% | 80% (random) | 20% (random) | Yes | 80% |
| 40% | 60% | 80% (1-35,101) | 20% (35,102-43,877) | No | 2.55% |
| 40% | 60% | 80% (6,777-0) | 20% | No | 7.9% |

Table 5. The results of different experiments

*4.2.1 XAI System Design.* The system flow for XAI begins with utilizing our Malware API dataset, which is then filtered to create a balanced dataset with 1079 malware and 1079 benign samples to maintain a 1:1 ratio. The balanced dataset undergoes pre-processing, including dropping the MD5 sum column and performing normalization, feature extraction, and transformation. Subsequently, multiple models (MLP, CNN, RNN, CNN-LSTM) are trained using the pre-processed data. LIME (Local Interpretable Model-agnostic Explanations) and SHAP (Shapley Additive Explanations) are applied to each trained model to understand and interpret the predictions by providing feature contribution insights. The results from LIME and SHAP generate visual interpretations of the model's predictions, which are then analyzed to derive meaningful insights. This structured approach ensures clarity, explainability, and transparency throughout the process, facilitating a deeper understanding of the models' decision-making. The system flow is illustrated in Fig. 6.

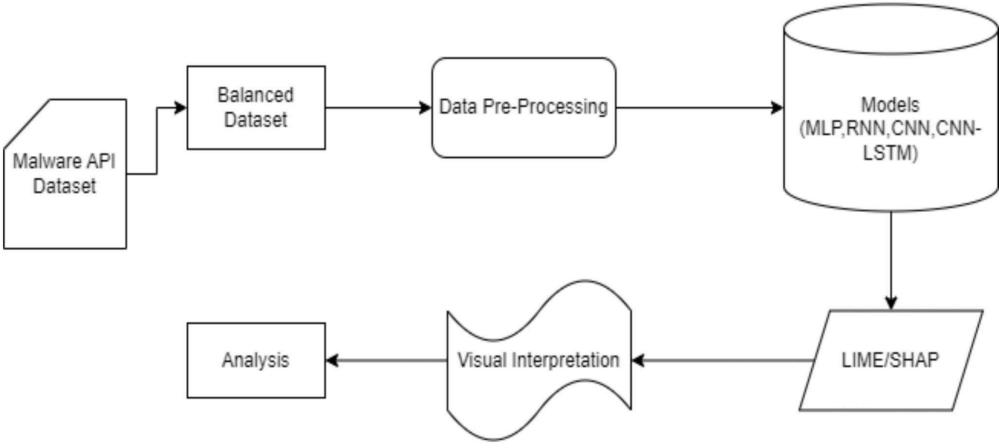

Fig. 6. XAI System Flow

4.3 Data-Set

We consider the balanced dataset described in section 3.1 to follow best practices, as concluded in the first section of our paper.

4.4 Operational Experiments

In this section, we evaluate all our models using traditional evaluation metrics. These metrics provide a comprehensive understanding of model performance across various dimensions. The visual interpretations, including the confusion matrix, loss curve, accuracy curve, and precision-recall curve, are supplemented by numerical metrics such as accuracy, macro average recall, and macro average precision. All metrics are listed in Table 7. We experimented with the models twice. Once we considered the whole dataset, we found that the dataset had 41,797 malware and 1,079 goodware. There is a visible imbalance in the classification, but during our literature review, we found many models claiming their accuracy directly in the dataset, which we found was biased. As stated, we observed that the data set is one of the factors influencing a model's performance, which is what we found in our experiment. In the first experiment, we took the complete dataset; in the other, we took the balanced dataset. We wanted to find out which model has the least influence on the change in the dataset and how the behavior of models changes.

| | | | | |
|---|---|---|---|---|
| Accuracy | 0.99 | 0.99 | 0.91 | 0.99 |
| Macro Average Recall | 0.74 | 0.92 | 0.87 | 0.86 |
| Macro Average Precision | 0.87 | 0.84 | 0.95 | 0.94 |
| Macro Average F1 Score | 0.79 | 0.88 | 0.91 | 0.89 |
| Weighted Average Recall | 0.98 | 0.99 | 0.99 | 0.99 |
| Weighted Average Precision | 0.98 | 0.99 | 0.99 | 0.99 |
| Weighted Average F1 Score | 0.98 | 0.99 | 0.99 | 0.99 |

Table 6. Summary of Evaluation Metrics for the entire dataset.

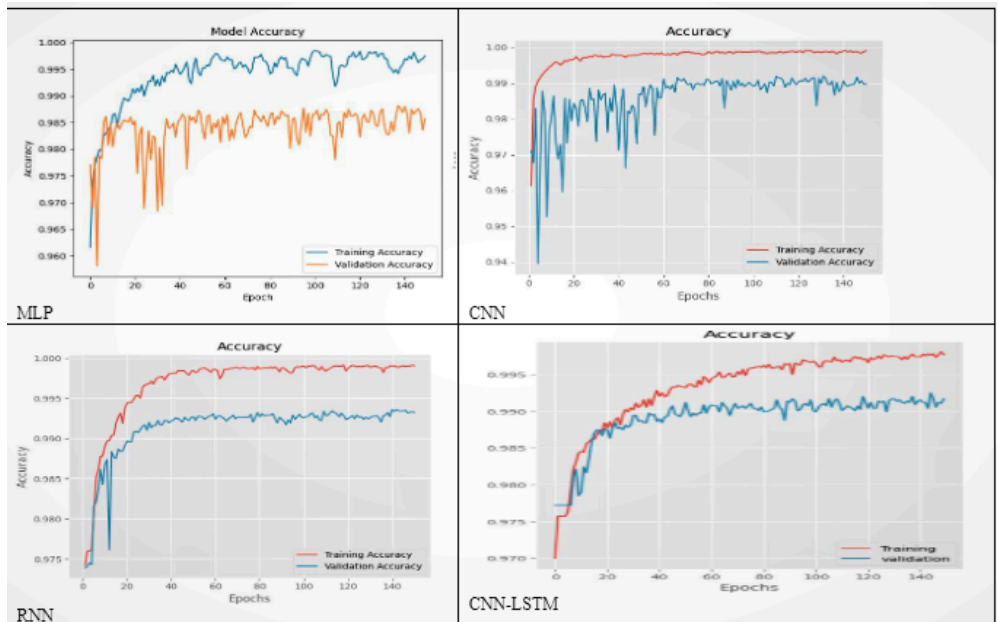

Fig. 7. Accuracy Vs. Epoch graph for the unbalanced dataset.

- With the entire dataset: On running the models, each for an epoch of 150, we found the output as shown in Table 6.
- Balanced dataset: On running our models each for 150 epochs, we found the output shown in Table 7.

| Metric | MLP | CNN | RNN | CNN-LSTM |
|---|---|---|---|---|
| Accuracy | 0.99 | 0.99 | 0.91 | 0.99 |
| Macro Average Recall | 0.74 | 0.92 | 0.87 | 0.86 |
| Macro Average Precision | 0.87 | 0.84 | 0.95 | 0.94 |
| Macro Average F1 Score | 0.79 | 0.88 | 0.91 | 0.89 |
| Weighted Average Recall | 0.98 | 0.99 | 0.99 | 0.99 |
| Weighted Average Precision | 0.98 | 0.99 | 0.99 | 0.99 |
| Weighted Average F1 Score | 0.98 | 0.99 | 0.99 | 0.99 |

Table 7. Summary of Evaluation Metrics with Balanced dataset for MLP, CNN, RNN, and CNN

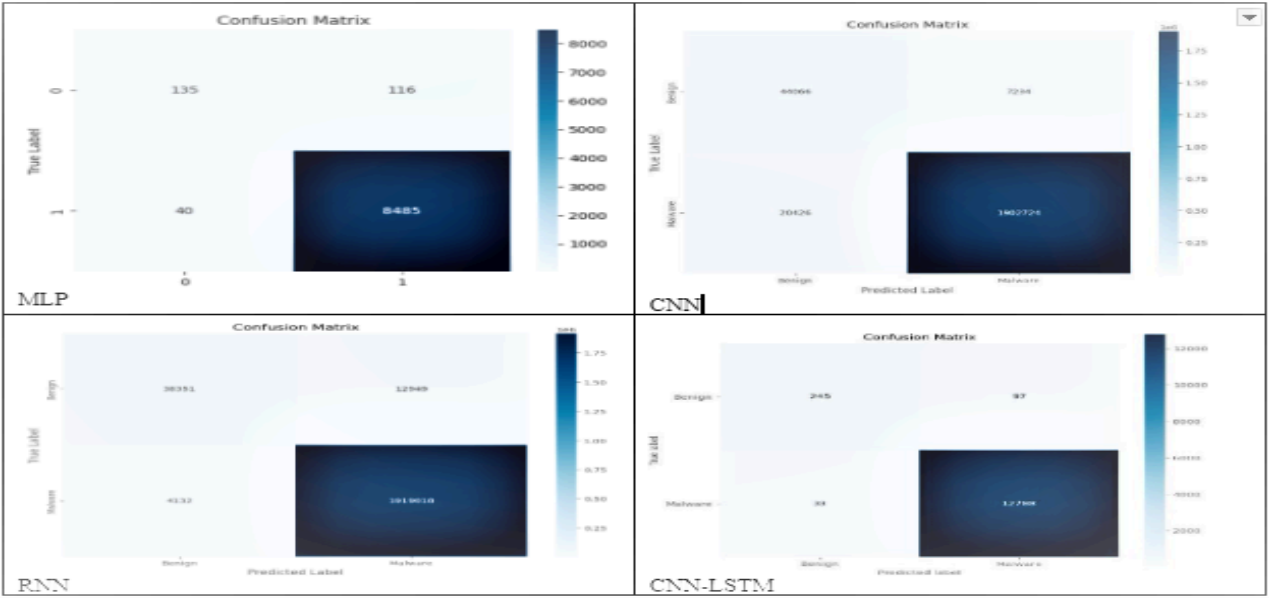

Fig. 8. Confusion Matrix for Models used based on Unbalanced Dataset.

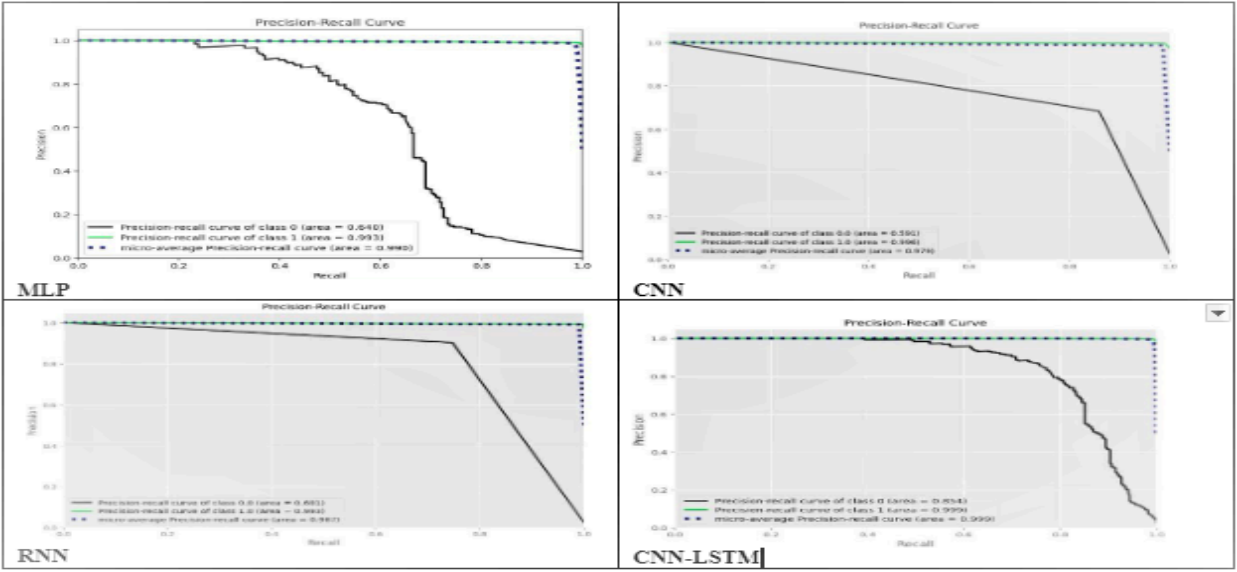

Fig. 9. Precision-Recall Curve Graph for an unbalanced dataset.

4.5 Analysis

In the case of an unbalanced dataset, the result is presented in Table 6. Also, the precision recall curve for the unbalanced dataset, has been plotted through graphs for the 4 models: MLP, RNN, CNN, and CNN-LSTM have been shown in Fig. 9. Similarly, the confusion matrix for the unbalanced dataset, has been plotted through graphs for the 4 models: MLP, RNN, CNN, and CNN-LSTM have been shown in Fig. 8. On the similar terms, Fig. 11, Fig. 12, Fig. 13, and Fig. 14 shows the accuracy/epoch graphs, confusion matrix, precision-recall graphs, and the ROC curve for the balanced datasets. The MLP accuracy curve shows steady improvement in training accuracy, but the validation accuracy fluctuates significantly, indicating some instability and potential overfitting(shown in Fig. 7). In CNN, both training and validation accuracy curves rise steadily with minor fluctuations, though the gap between the two curves suggests slight overfitting. In RNN, the training accuracy improves over time, but the validation accuracy shows high variability, indicating unstable generalization and potential overfitting. CNN-LSTM: The accuracy curves for both training and validation(Fig. 7) are smooth and closely aligned, showing strong learning and good generalization with minimal overfitting. In MLP, the ROC curve(Fig. 10) shows a good separation with a large area under the curve (AUC), indicating that the model performs well in distinguishing between classes with high sensitivity and specificity. The CNN ROC curve (shown in Fig. 10) is also strong but slightly less smooth than the MLP, with a high

AUC, suggesting good classification performance, though with some variability. RNN: The RNN's ROC curve shows some inconsistency, indicating that its ability to distinguish between classes is weaker than MLP and CNN, leading to a lower AUC.

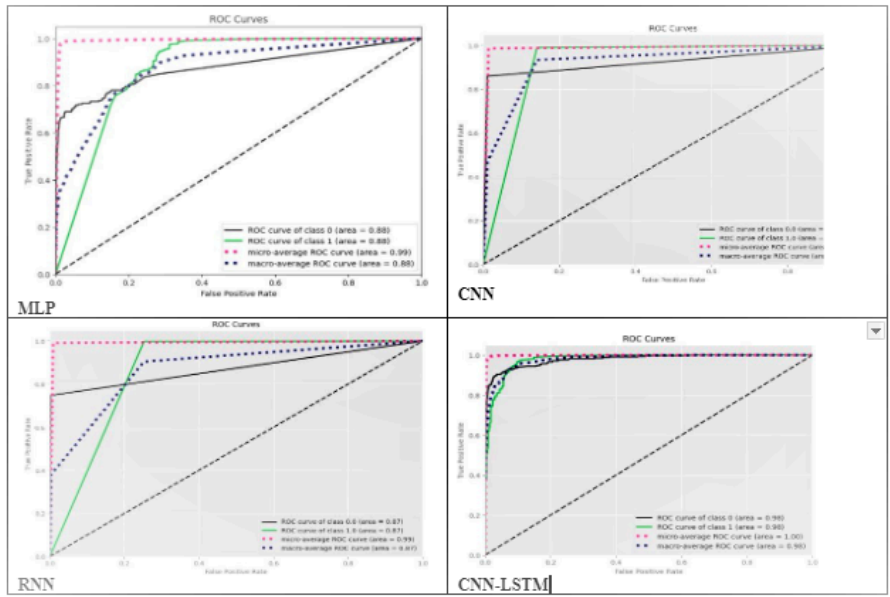

Fig. 10. ROC Curve for Models based on Unbalanced Dataset

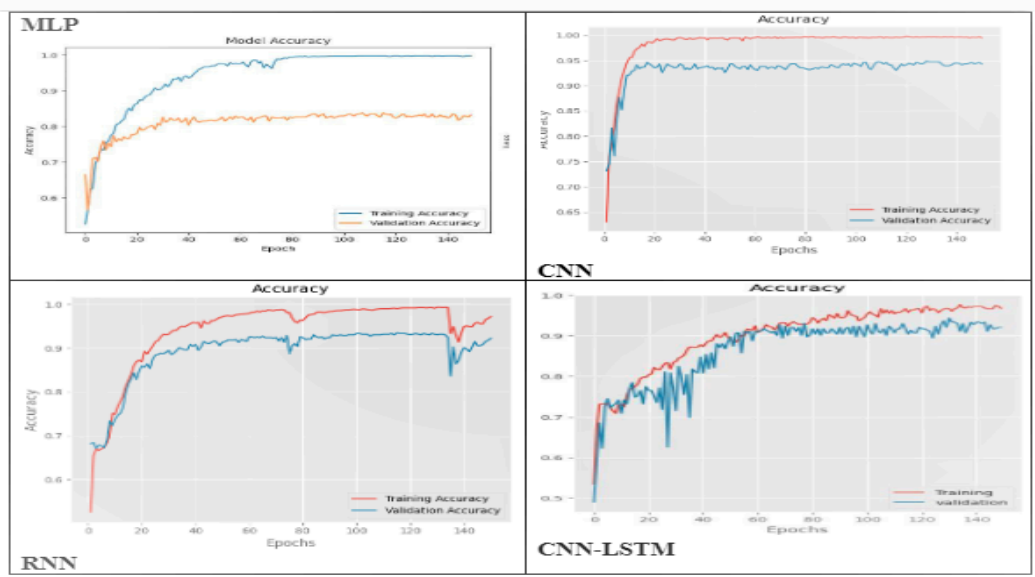

Fig. 11. Accuracy vs epoch for different models for the balanced dataset based on the values of Table 7

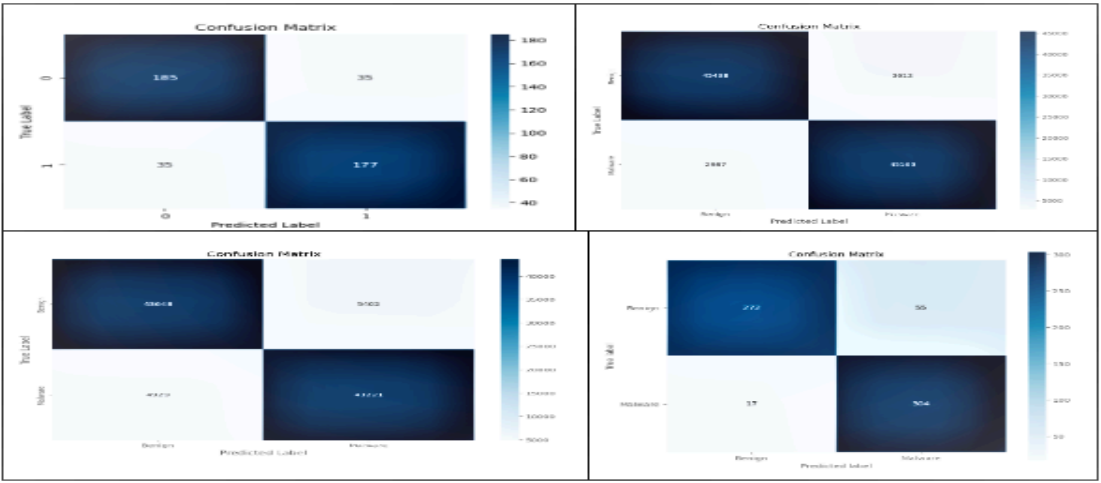

Fig. 12. Confusion Matrix for different Models for the Balance dataset based on the values of Table 7.

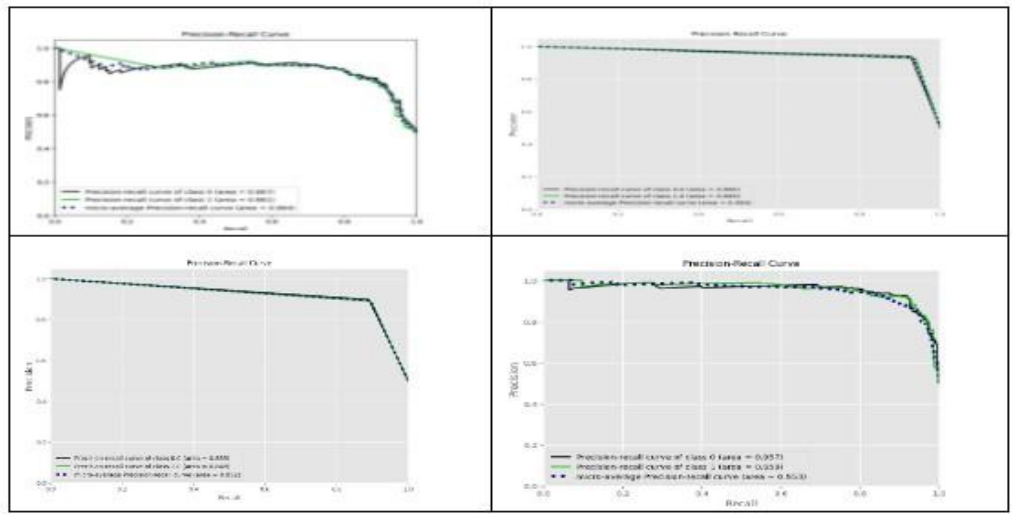

Fig. 13. Precision-recall curve for different Models for the Balance dataset based on the values of Table 7.

The ROC curve for CNN-LSTM(Fig. 10) is smooth and closely follows the top-left corner, representing the best performance with high AUC, indicating excellent classification ability and robustness in handling both classes, as can be seen in Table 6. Among models, CNN showed the smallest accuracy difference between unbalanced and balanced datasets, demonstrating robustness to class imbalance, whereas MLP struggled the most. In the case of the balanced dataset, the result is presented in Table 7, and the analysis reveals that MLP training accuracy improves steadily. Still, the validation accuracy is much lower and fluctuates, indicating difficulty in generalizing to the balanced dataset. In CNN, the accuracy curves for both training and validation show a close alignment with slight fluctuations, indicating stable learning and good generalization performance on the balanced dataset. The RNN accuracy curve shows a slower and more unstable rise in validation accuracy, with significant fluctuations, suggesting that the model struggles with consistent learning and generalization. The CNN-LSTM shows a smooth and steady improvement in training and validation accuracy, with both curves closely aligned, indicating strong generalization and effective learning on the balanced dataset. The ROC curve has a high AUC, meaning the model correctly distinguishes between positive and negative classes, with strong sensitivity (true positive rate) and low false positives in MLP. The CNN ROC curve stays close to the top-left corner with a very high AUC, indicating excellent performance in distinguishing between classes, with a good balance between true positives and false positives. The ROC curve for the RNN shows some fluctuations, with a slightly lower AUC, indicating the model is less consistent at distinguishing between positive and negative classes compared to MLP and CNN. The ROC curve is smooth and has a high AUC, indicating strong performance in distinguishing between positive and negative classes, similar to CNN, with good sensitivity and low false positives in CNN-LSTM. The CNN model shows the smallest difference in accuracy between unbalanced and balanced datasets, indicating its robustness and generalization capability. The highest accuracy difference (0.15) in MLP shows its struggle with class imbalance. This analysis concludes that while class imbalance can significantly affect model performance, models like CNN can generalize well across varied datasets. In contrast, models like MLP may require further tuning to handle such imbalances effectively. It was found that the accuracy of the models decreased when they were operated on the balanced dataset. This shows that data sets affect models' performance, so they must be curated with immense attention to replicate real-world scenarios.

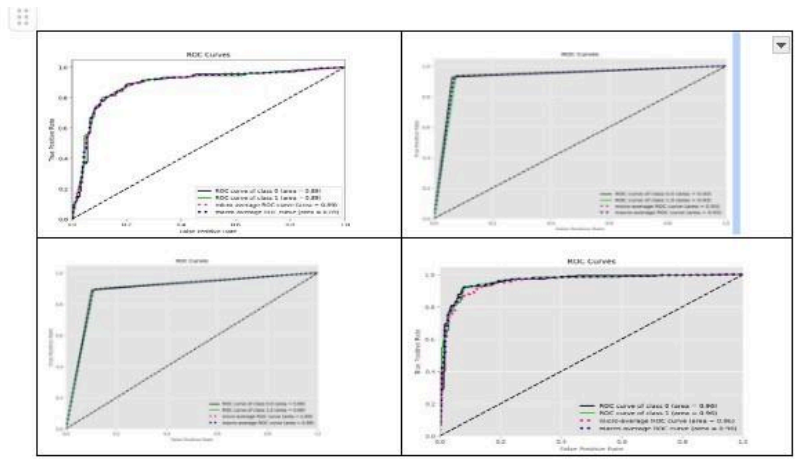

Fig. 14. ROC curve for a balanced dataset for all the models

4.6 Explainable AI Experiments

4.7 LIME Analysis

In this step, we run LIME; the basic working of LIME is explained in Section 3.2.5 and has been applied to all four deep-learning models to gain insights into which features had more effect on the performance of the model by considering accuracy. In evaluating the LIME outputs for various models, two primary metrics were considered: prediction probabilities and feature impacts (shown in Fig. 15). The analysis focused on these aspects across four models: MLP, RNN, CNN, and CNN-LSTM. The CNN model exhibited high confidence, especially in identifying non-malware, with a prediction probability of 0.35 for Non-Malware. The MLP model, with a Malware probability of 0.36 and Non-Malware prediction probability of .64, shows a high sensitivity to features influencing Non-Malware detection, suggesting usefulness in contexts where detection of legitimate packets is a priority and in situations where attackers use decoy techniques to infiltrate(Fig. 15). The RNN has a prediction probability of 0.99, making it ideal for scenarios demanding minimal false positives. Therefore, it could be challenging when intelligent devices are targeted with adversarial or data poisoning attacks on ML models. The RNN and CNN-LSTM models displayed more balanced predictions between Malware and Non-Malware categories, indicating potential robustness and better generalization capabilities. Feature t.71 has the highest impact, with a value of 2.96, pushing the prediction towards Malware. Feature t.60 also positively contributes to predicting Malware, with a value of 1.90. In contrast, Feature t.49 contributes negatively with a value of -1.73, pushing the prediction towards Non-Malware.

In the case of CNN, t.52 has a high value of 297, strongly contributing to the Malware classification. t.3, with a value of 16.60, pushes the prediction towards Non-Malware.

In the case of RNN, t_40 (value 112.00) and t_44 (value 158.00) are significant features influencing the prediction towards Non-Malware. t_84 (value 40.00) slightly contributes towards Malware, but has much less influence than Non-Malware pushing features. The strong confidence in the Non-Malware classification comes from multiple features with high values (such as t_40, t_44, and t_75) pushing the decision towards Non-Malware. Only a few features, like t_84, influence a malware prediction, but their impact is minimal compared to those driving the Non-Malware prediction.

Finally, with CNN-LSTM, t_52 (value 297.00) has the highest impact, strongly pushing the prediction towards Non-Malware. t_46 (value 274.00) and t_75 (value 226.00) also significantly contribute to classifying the input as Non-Malware. In the feature value graph, the prediction is a close call between

Malware and Non-Malware, but the Non-Malware class has a slight edge due to the stronger influence of features like t_52, t_46, and t_75. Features like t_69 attempt to push the prediction toward Malware, but don't outweigh the stronger features pushing toward Non-Malware (shown in Fig. 16).

Conclusively, while the CNN model scores high on accuracy for non-malware identification, the MLP might be preferable for its broad sensitivity to malware-indicative features, making model selection context-dependent.

4.8 SHAP Analysis

After applying SHAP on all four models, we analyze each model by visualizing three plots. The images depict different visualizations created using SHAP (Shapley Additive explanations), a powerful tool for interpreting machine learning models. These visualizations are essential for understanding, diagnosing, and communicating the behavior of complex models in a comprehensible way and help explain a model's output by showing the impact of each feature on the model's prediction as shown in Fig. 17 and Fig. 18. The key visualization plot used is the Waterfall plot: this helps to know how each feature contributes to shifting the model output from the base value (the average model output over the dataset) to the actual prediction for a specific instance.

Summary Plot: This plot helps identify which features are most important for the model across the entire dataset and illustrates how high or low feature values affect the predictions. Bar Plot: provides a clear view of which features had the most significant impact, positively or negatively, on that specific outcome. Key insights regarding their performance and feature impact can be drawn based on the SHAP analysis visualizations for the MLP, CNN, RNN, and CNN-LSTM models. The MLP (Multi-Layer Perceptron) model demonstrates a balanced feature impact across its predictions. The SHAP Waterfall model for MLP shows that the features incrementally push the prediction towards or away from the final class (Non-Malware or Malware), with t_52 contributing strongly towards Non-Malware. Like MLP, the CNN model shows key features pushing the prediction towards Malware, with t_52 and t_75 being critical contributors. The RNN shows a sharp increase for Malware, but t_44 strongly influences Non-Malware. For CNN-LSTM, the feature t_52 significantly drives the prediction towards Non-Malware, with features like t_15 attempting to pull the prediction towards Malware.

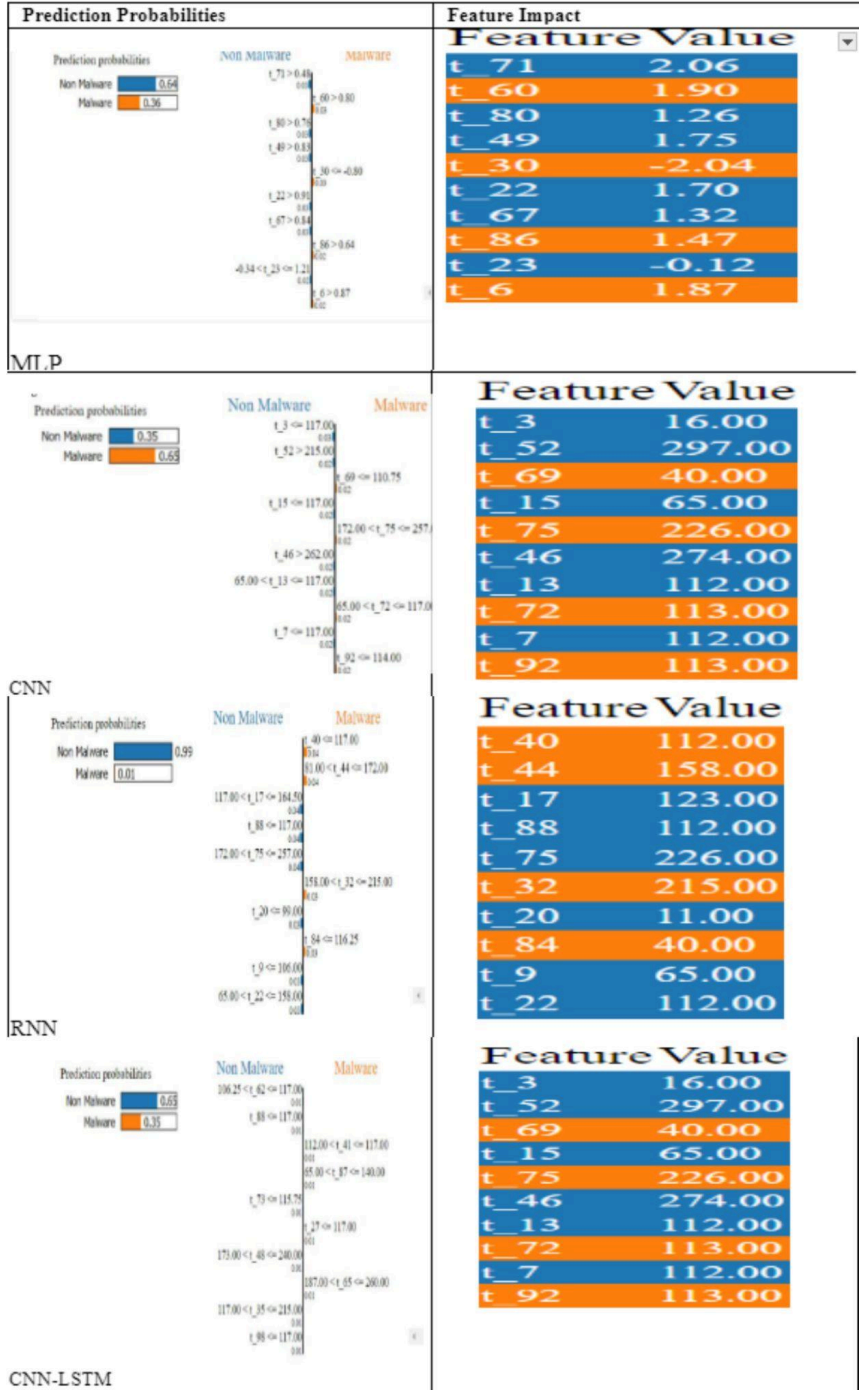

Fig. 15. LIME Analysis Visualizations

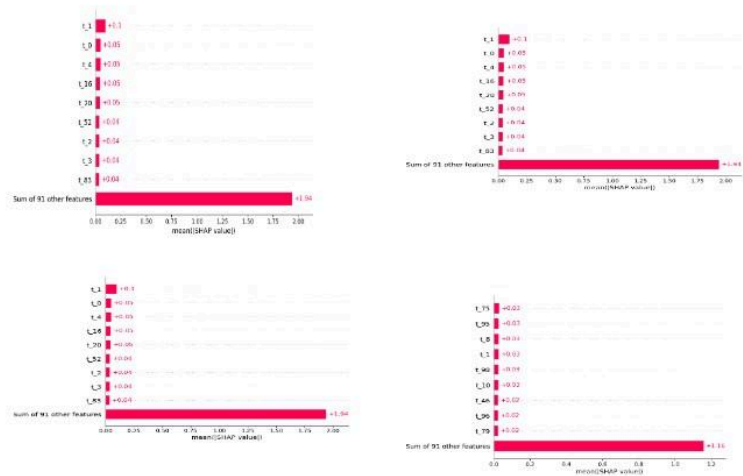

Fig. 16. Feature-value graph

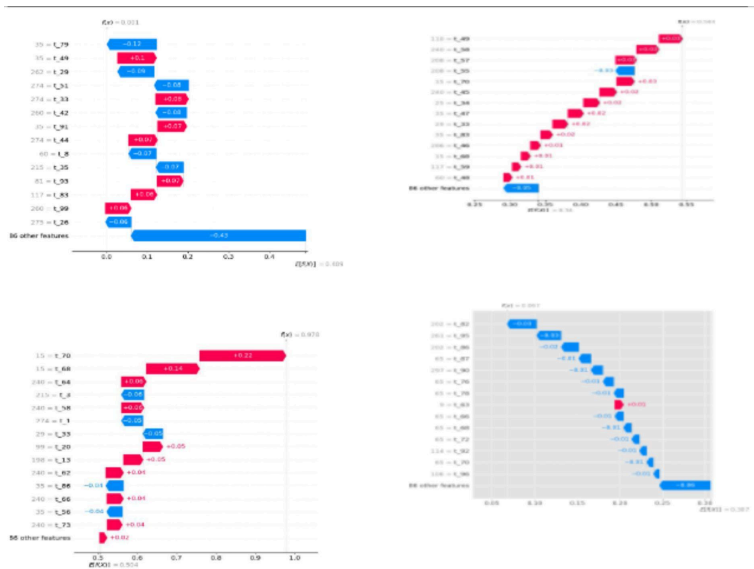

Fig. 17. SHAP analysis visualization

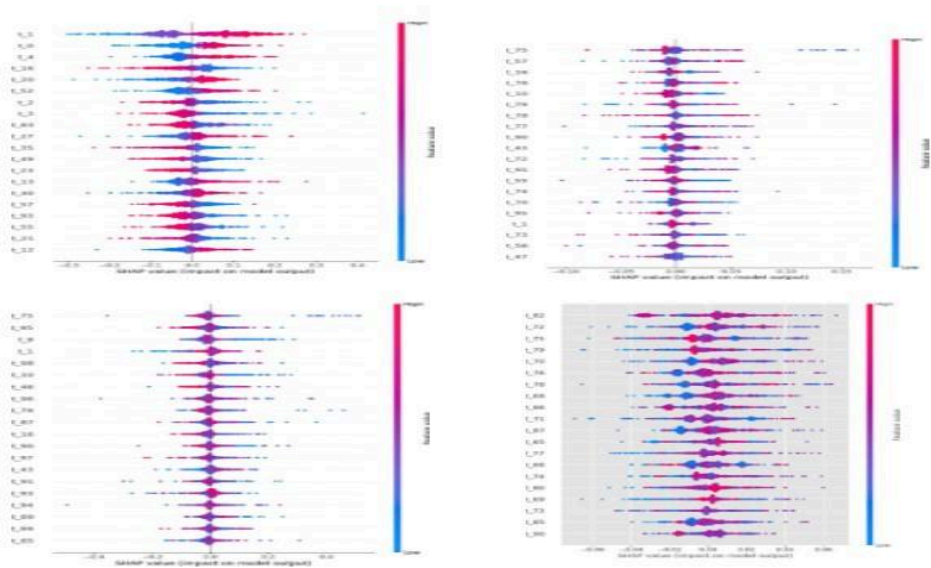

Fig. 18. SHAP analysis for feature value

**5 Conclusion**

In this paper, we presented a novel approach for malware detection and analysis using Explainable AI in two parts. In the first part, we analyzed the validity of using an MLP (Multilayer Perceptron) for malware detection. In the second part, we evaluated four models (MLP, CNN, RNN, and CNN-LSTM) and provided explainability to these models to increase the trust of developers. Our approach aimed at effectively differentiating between malicious and benign software samples.

Our primary strategy revolved around the implementation of an MLP. In addition to the MLP, we also examined the capabilities of Support Vector Machine (SVM), Convolutional Neural Network (CNN), and Recurrent Neural Network (RNN) models. The MLP model demonstrated superior accuracy compared to the others. Experiments with the MLP model revealed a consistent level of performance, which improved as the complexity of the model increased. The most effective configuration was a three-layer structure consisting of 100 neurons. Our analysis focused on API call sequences, reflecting the dynamic nature of malware interactions with the operating system and offering insightful runtime data. To facilitate a balanced evaluation, we created a standardized testing environment for the SVM, CNN, and RNN models. Notably, our MLP model consistently surpassed the performance of these models, maintaining an accuracy rate of 0.98. Our analysis showed that the MLP model gave an accuracy of 0.78 on a balanced dataset. However, overfitting occurred when using SMOTE to balance the dataset. By incorporating epochs in our training, we overcame this overfitting and achieved an accuracy of 0.83.

In the second part of our research, we analyzed the deep learning models (MLP, CNN, RNN, and CNN-LSTM) using LIME and SHAP to understand our models' behavior. Using LIME, we identified significant influences on the predictions. The SHAP analysis provided several key insights:

• The MLP model exhibited a balanced distribution of feature impacts, with a wide spread of SHAP values indicating an influence from multiple features.
• The CNN model showed a concentrated feature impact, with specific features strongly influencing predictions, enhancing performance in tasks where these attributes are highly predictive.
• The RNN model had a diverse distribution of SHAP values, with some features having more significant impacts. • The CNN-LSTM model revealed a distinct pattern with a strong emphasis on a few features, indicating its effectiveness in leveraging key attributes for predictions.

Conclusively, while the CNN model scores high on accuracy for non-malware identification, the MLP might be preferable for its broad sensitivity to malware-indicative features, making model selection context-dependent. The SHAP analysis highlighted the importance of feature selection and its impact on the performance of these models, with CNN-LSTM and CNN standing out in their ability to utilize crucial features effectively.

## 6 Limitations

It's essential to note that the scope of our paper did not include addressing intelligent malware capable of deceiving sandboxing environments, which can be considered future work for the research community. The robust and efficient performance of our MLP model, particularly in the analysis of API call sequences, positions it as a potent and adaptable tool for detecting malware across diverse settings. This study contributes valuable insights to the field of malware detection, paving the way for further advancements in cybersecurity.